\documentclass[sigconf,authorversion]{acmart}

\usepackage{subcaption}
\usepackage[ruled]{algorithm2e} 
\SetAlFnt{\scriptsize}
\SetAlCapFnt{\small}
\SetAlCapNameFnt{\small}
\SetAlCapHSkip{0pt}

\usepackage{xcolor}

\copyrightyear{2022} 
\acmYear{2022} 
\setcopyright{acmlicensed}\acmConference[SA '22 Technical Communications]{SIGGRAPH Asia 2022 Technical Communications}{December 6--9, 2022}{Daegu, Republic of Korea}
\acmBooktitle{SIGGRAPH Asia 2022 Technical Communications (SA '22 Technical Communications), December 6--9, 2022, Daegu, Republic of Korea}
\acmPrice{15.00}
\acmDOI{10.1145/3550340.3564223}
\acmISBN{978-1-4503-9465-9/22/12}

\setcitestyle{square}

\begin{document}

\title[Progressive Material Caching]{Progressive Material Caching}

\author{Shin Fujieda}
\affiliation{%
  \institution{Advanced Micro Devices, Inc.}
  \country{Japan}}
\email{Shin.Fujieda@amd.com}
\orcid{0000-0002-2472-7365}

\author{Takahiro Harada}
\affiliation{%
  \institution{Advanced Micro Devices, Inc.}
  \country{USA}}
\email{Takahiro.Harada@amd.com}
\orcid{0000-0001-5158-8455}




\renewcommand{\shortauthors}{Fujieda and Harada}

\begin{abstract}
The evaluation of material networks is a relatively resource-intensive process in the rendering pipeline. Modern production scenes can contain hundreds or thousands of complex materials with massive networks, so there is a great demand for an efficient way of handling material networks.
In this paper, we introduce an efficient method for progressively caching the material nodes without an overhead on the rendering performance.
We evaluate the material networks as usual in the rendering process. Then, the output value of part of the network is stored in a cache and can be used in the evaluation of the next materials. Using our method, we can render the scene with performance equal to or better than that of the method without caching, with a slight difference in the images rendered with caching and without it.
\end{abstract}




\begin{teaserfigure}
    \begin{subfigure}[b]{0.245\textwidth}
        \centering
        \includegraphics[width=\textwidth]{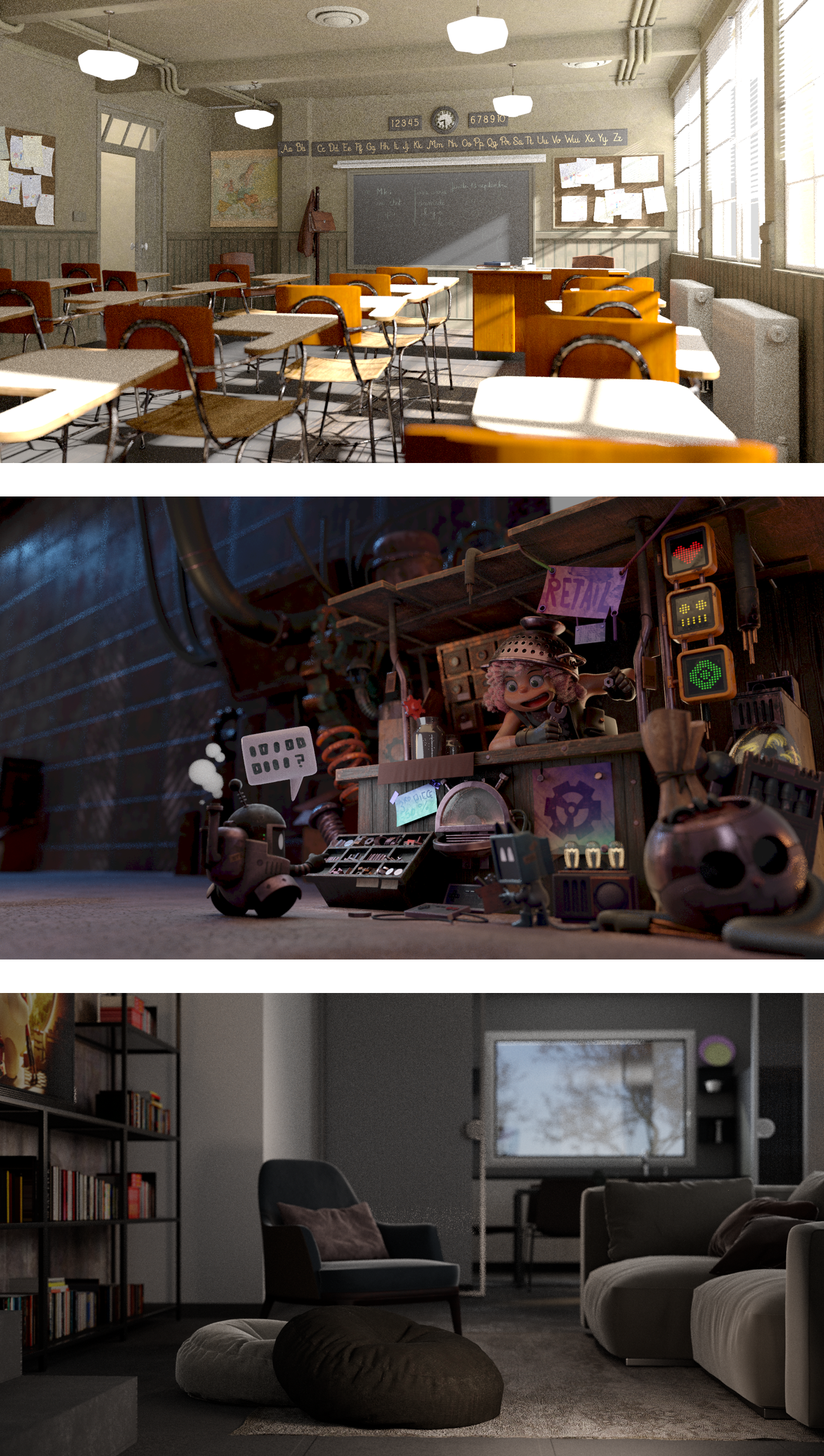}
        \caption{Without caching}
    \end{subfigure}
    \hfill
    \begin{subfigure}[b]{0.245\textwidth}
        \centering
        \includegraphics[width=\textwidth]{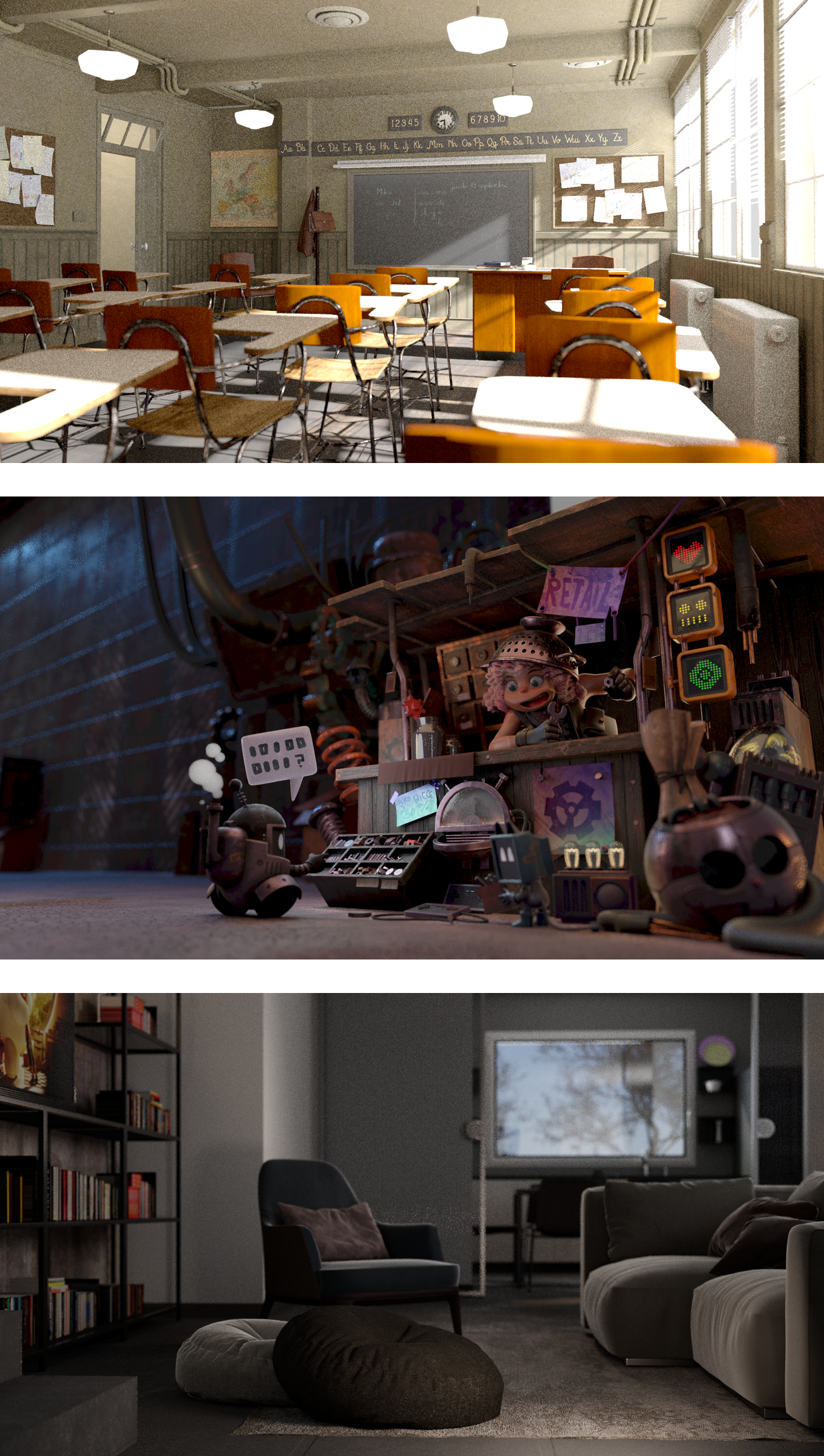}
        \caption{Our method}
    \end{subfigure}
    \hfill
    \begin{subfigure}[b]{0.245\textwidth}
        \centering
        \includegraphics[width=\textwidth]{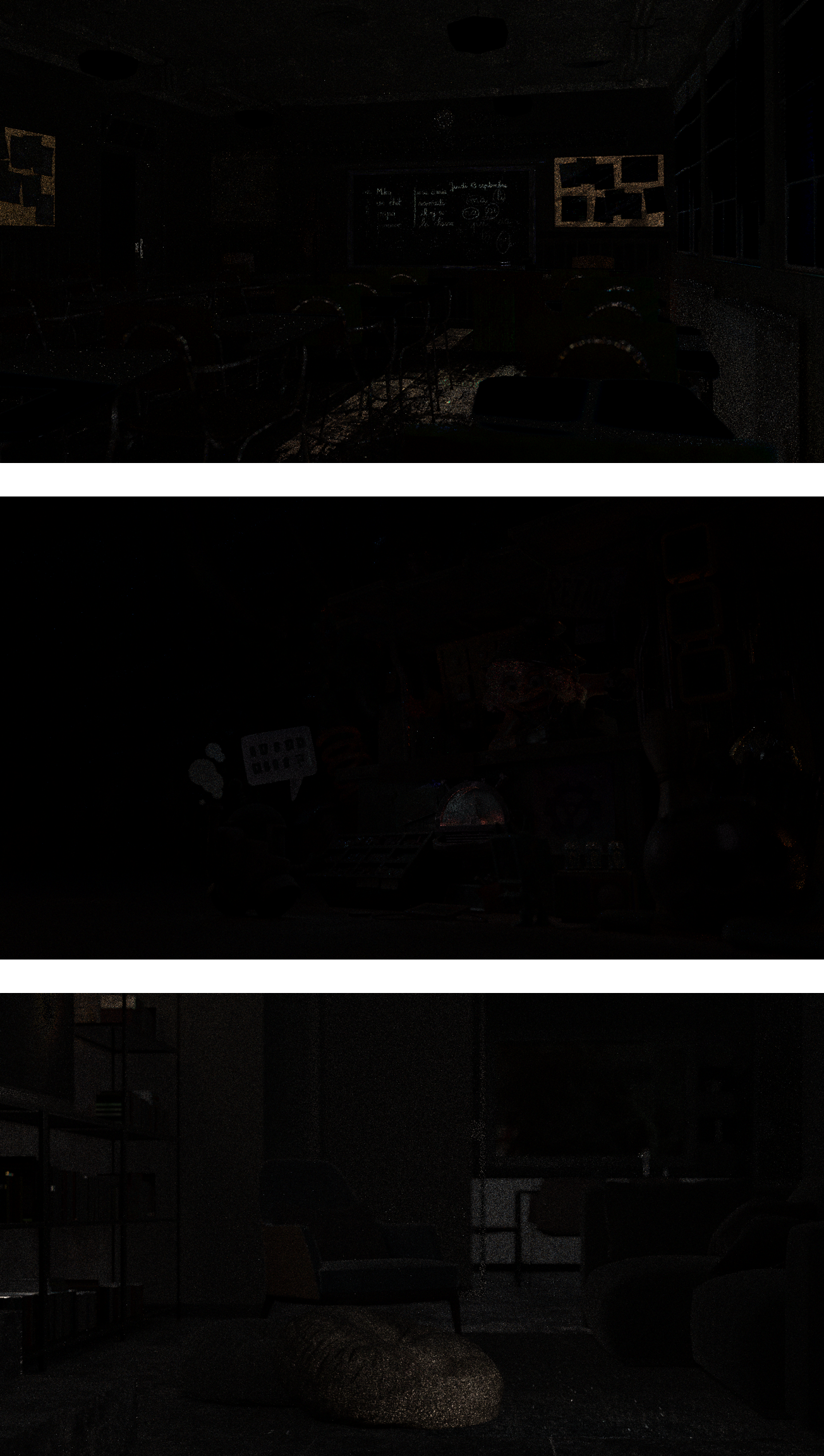}
        \caption{Difference between (a) and (b)}
    \end{subfigure}
    \hfill
    \begin{subfigure}[b]{0.245\textwidth}
        \centering
        \includegraphics[width=\textwidth]{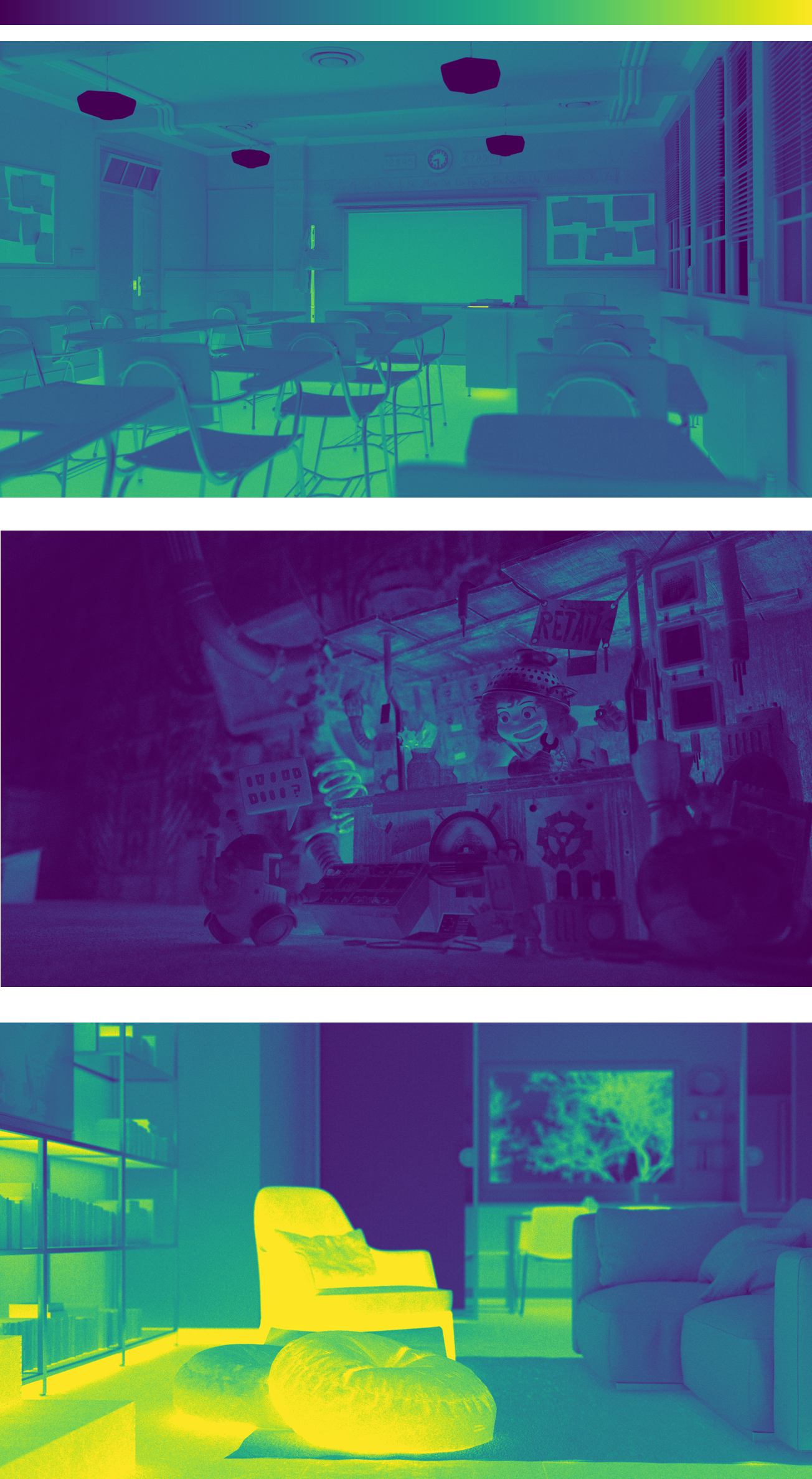}
        \caption{Nodes found in the cache}
    \end{subfigure}
    \caption{Equal-sample comparisons (128 samples per pixel) with {\sc Classroom}, {\sc Junk Shop} and {\sc Italian Flat} scenes. 
    (a) Images rendered without caching. (b) Images rendered with our method. (c) Differences between images (a) and (b) multiplied by 5. (d) Visualizations of the average number of material nodes per sample which can be found in the cache by lookup. As the color changes from black to yellow in the viridis colormap, the respective average increases from 0 to 20.}
    \label{fig:teaser}
\end{teaserfigure}
\maketitle
\section{Introduction}
Modern 3D modeling software allows artists to create their desired scenes using node-based materials, which make editing these materials more intuitive and flexible.
Thanks to this node-based system, most scenes in production rendering usually contain hundreds or thousands of materials with large material networks.
Rendering these scenes requires us to evaluate complex materials, which is generally a very resource-intensive operation that often can be the bottleneck of the rendering performance.
Besides, we need to evaluate the same complex network many times, which is redundant and can be optimized.
To alleviate this bottleneck and redundancy, one can bake all inputs of BSDF nodes to texture maps~\cite{substance2022} or per-vertex data~\cite{Manuka2018} as a pre-process and use them when evaluating the materials in the rendering process.
However, this pre-process requires some additional resources and cannot handle everything properly because some material nodes require information about the object at a shading point.

In this paper, we propose a method to progressively cache outputs of the material nodes instead of pre-processing them.
We call this a material cache which is queried by a combination of several keys considering shading-point information using a hash table (see Sec.~\ref{pmc}) as we render the scene.
Then, we can reuse the cache when a ray hits the surface with the material under the same condition next time.
We also present our GPU implementation that inserts the output value of a material node into a fixed-size cache with two different hashes.
This progressive way of handling the material cache does not add any overhead on the rendering performance, except for the additional memory for the cache.
To summarize, the contributions of our work are as follows:

\begin{itemize} 
\item We introduce a progressive method to cache some material nodes without an overhead on the rendering performance.
\item We show an efficient approach to handling the UV value to store material nodes in the cache.
\item This paper also illustrates the implementation details of our method to update a hash table on the GPU.
\end{itemize}

\section{Material Network Evaluation} \label{sec2}

\begin{figure}[t]
  \centering
  \includegraphics[width=\linewidth]{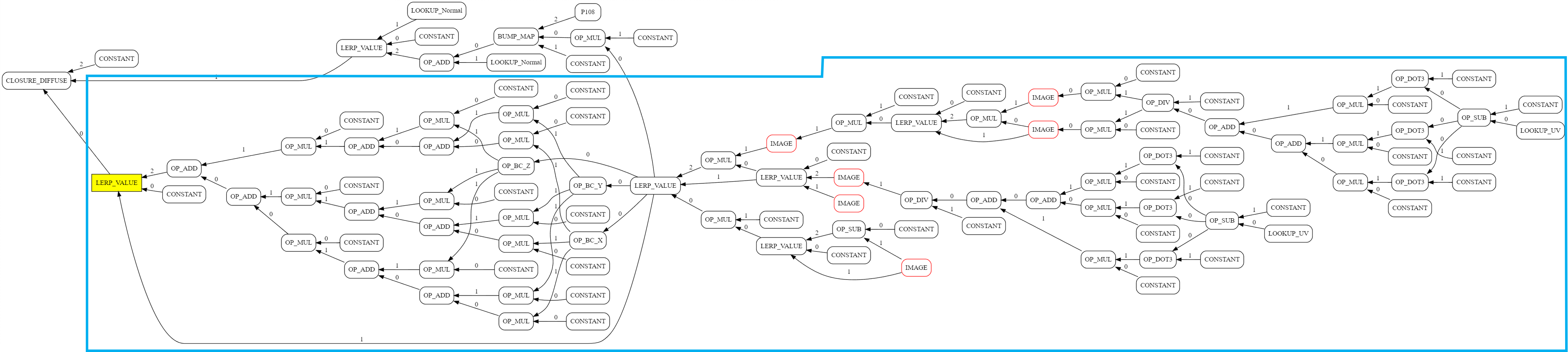}
  \caption{\label{fig:matExample}
           An example of a material. The yellow and red nodes are the cacheable and UV-dependent nodes, respectively.}
\end{figure}

Fig.~\ref{fig:matExample} shows one of the material networks in the {\sc Classroom} scene. Larger image can be found in our supplemental material.
This material is composed of a massive network having more than 100 nodes.
To render the scene, we need to evaluate such massive networks many times, so efficiently handling them is crucial to the rendering performance.

In our implementation, we process the material networks as a form of stack machine on the GPU.
More specifically, nodes in a material graph are converted into data that is read and executed in the order they are stored in the memory.
Although there are other ways of processing the material networks on the GPU, such as a dynamic kernel generation per material, we choose it because we need to keep the kernel compilation time manageable, even if the scene has a lot of materials.
Also, we optimize the material networks before rendering the scene by converting them into forms that can be easily placed in and treated by the stack, for example, by pre-computing simple arithmetic nodes with constant inputs.
\section{Progressive Material Caching} \label{pmc}
Our method progressively caches the outputs of some material nodes as we render the scene with little pre-process of material networks.
Before rendering the scene, we traverse the material networks from the bottom to the top to define cacheable nodes at the highest possible level in the network.
A cacheable node, where a result of its sub-graph is cached, is a node whose sub-graph has no shading-point dependent node as a descendant other than texture coordinates (UVs), as we are going to describe in Sec.~\ref{descriptor}.
And then, when reaching the cacheable node during a material evaluation in the rendering process, we first look it up in the cache, and, if it is found, we can directly use the cached value without processing its sub-graph under the cacheable node.
Otherwise, we manage its sub-graph in the standard way and store the resulting value in the cache.
Thus, by defining cacheable nodes at highest possible levels in a network, we can skip the material evaluation of most parts of this network.
Fig.~\ref{fig:matExample} shows an example of a material where a cacheable node is marked with yellow and its sub-graph is outlined with the blue polygon.
In this material, the cacheable node is defined at the highest level, so we do not need to process most parts of the network once it is cached.
We can have multiple cacheable nodes for a single material.
Note that this process of defining the cacheable nodes is a lightweight computation, because we can execute it at the same time as we optimize the material networks as described in Sec.~\ref{sec2}. 

\begin{figure}[t]
  \centering
  \includegraphics[width=0.8\linewidth]{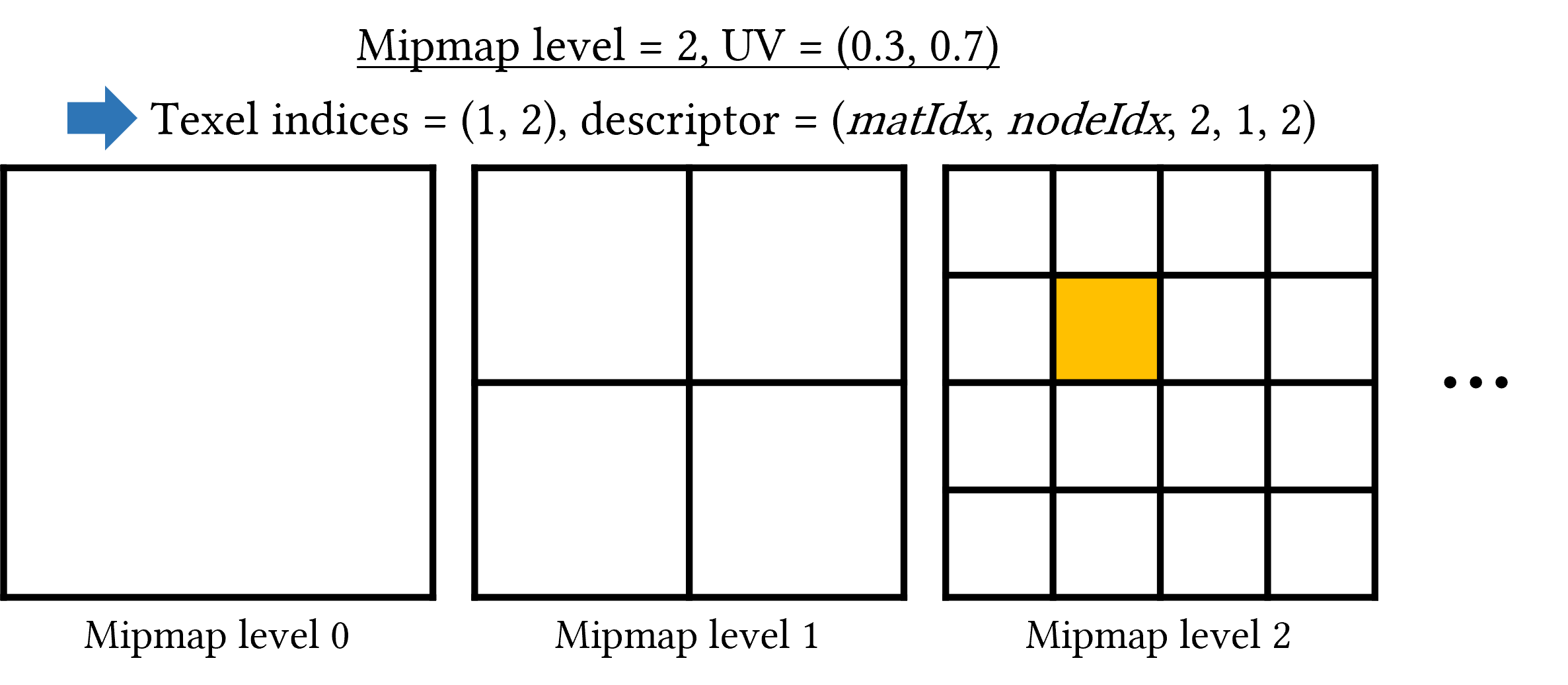}
  \caption{\label{fig:virtualMip}
           Virtual mipmapped texture.}
\end{figure}

\subsection{Node Descriptor with UV} \label{descriptor}
We need to uniquely identify cacheable nodes to provide for caching their output values.
For each cacheable node, we define a descriptor that contains at least identifiers such as \textit{material ID} and \textit{node ID} that is the index of the node in the material network.
Additionally, as some material nodes depend on information about the object at a shading point, the descriptor should include it as well, to be unique depending on the shading point.
In this paper, we focus on UV, although other shading-point information, such as a world-space position, could also be used.
We do not consider a node cacheable if its sub-graph contains any shading-point dependent nodes other than UV.
We want to extend our method to them in the future.

We employ the UV value as one of the descriptors of cacheable nodes because it is the value most frequently used to handle material nodes.
With UV, the output value of the cacheable node can be cached as a texel value.
However, the challenge lies in defining the proper cache resolution, which is not obvious only from the UV value.
To properly define it, we also need other information, such as texture resolution and the mipmap level.
However, we do not know it when managing the cacheable node during the material evaluation at a shading point.
As seen from Fig.~\ref{fig:matExample}, for example, the UV-dependent nodes are different from the cacheable node and only exist on its sub-graph, so it is impossible to retrieve the texture information at the cacheable node.
Therefore, instead of using the actual texture information, we propose to introduce a virtual mipmapped texture for the cacheable node using the footprint of the ray cone at a shading point~\cite{Akenine-Moller2021LOD}.
We use the word "virtual" for this texture because texel allocation is on-demand due to the sparse texel access, i.e., we do not need to allocate a series of two-dimensional textures but only allocate texels we need.
This allows us to cache the information at the appropriate resolution.
The intersection of a ray cone and a surface gives an elliptic footprint, and we can compute the texture-coordinate gradients, $\vec{g_1}$ and $\vec{g_2}$, along with both axes of the ellipse, as derived in ~\cite{Akenine-Moller2021LOD}.
With these gradients, we compute the mipmap level $\lambda$ for the virtual texture as follows:
\begin{equation}
    \lambda = -\log_2 (min(\|\vec{g_1}\|, \|\vec{g_2}\|)),
\end{equation}

Considering this mipmap level, we calculate the resolution of the virtual texture from $2^\lambda$.
So, unlike the standard mipmap hierarchy, the resolutions of virtual textures increase as the mipmap level becomes higher in our definition, as shown in Fig.~\ref{fig:virtualMip}.
Then, using the UV value, the texel indices on the virtual texture are calculated.
As a result, we store the output value as a texel value of the virtual texture in the cache using the mipmap level $\lambda$ and the texel indices as the descriptor of the cacheable node.

\begin{figure}[t]
  \centering
  \includegraphics[width=0.75\linewidth]{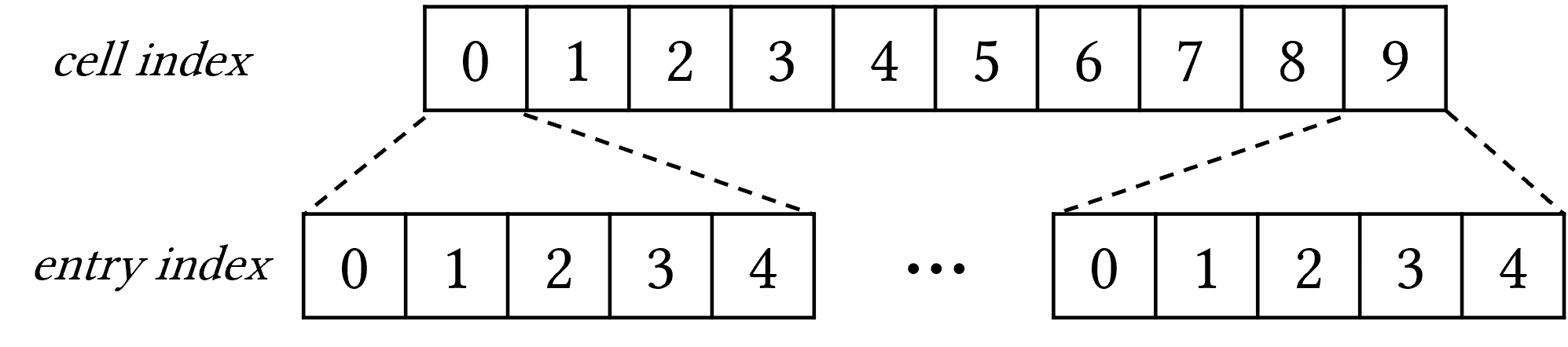}
  \caption{\label{fig:2hashes}
           $10\times5$ size of a hash table. The numbers of cells and entries on the hash table are $10$ and $5$ for each.}
\end{figure}


\begin{algorithm}[t]
\caption{Update of a hash table on the GPU}
\label{alg:update}
\SetKwInOut{Input}{Input}
\SetKwFunction{Hash}{Hash}
\SetKwFunction{HashSec}{Hash2}
\SetKwFunction{Encode}{Encode}
\SetKwFunction{CAS}{CAS}

\KwIn{Descriptor of the cacheable node $desc$, the cacheable node's value $v$, array of material cache $matCache$, the number of cells $N_c$ and the number of entries $N_e$.}
\BlankLine

$cellIdx \gets \Hash{desc} \% N_c$\;
$hashVal \gets \HashSec{desc}$\;
$hashIdx \gets -1$\;
\For{$i \leftarrow 0$ \KwTo $N_e$}
{
    $entryIdx \gets cellIdx \times N_e + i$\;
    $current \gets matCache[entryIdx]$\;
    \If{$current.hash = hashVal \cup current.hash = 0$}{
        $hashIdx \gets entryIdx$\;
        $break$\;
    }
}
\If{$hashIdx \neq -1$}
{
    $encodedVal \gets \Encode{v}$\;
    $new \gets \{hashVal,encodedVal\}$\;
    $old \gets \CAS{matCache[hashIdx], 0, new}$\;
}
\end{algorithm}


\subsection{Hashing Material Nodes} \label{hashing}
Using the unique descriptor discussed in Sec.~\ref{descriptor} as the key, we store and look up the cacheable node's value in a hash table to cache it in constant time.
Although the number of values of the cacheable nodes stored in the hash table is unknown because we store them progressively, we use a fixed-size hash table for the GPU efficiency.
As the size of the hash table is limited, hash collisions may occur, because different descriptors may be mapped to the same index in the table.
To differentiate each descriptor, we apply two different hash functions to the same descriptor as described in~\cite{Binder2022}.
In our implementation, the first hash function defines the \textit{cell index} on the fixed-size hash table with a modulo operator, whereas the second hash function checks the equality of keys and allocates the entry on each cell with the \textit{entry index}, as shown in Fig.~\ref{fig:2hashes}.
These numbers of cells and entries are chosen empirically.
Then, during the rendering, we progressively insert the cacheable node's value into the hash table using these two different hash values per the descriptor of the cacheable node.
However, even with this approach, we cannot completely avoid hash collisions.
In this case, we keep the first inserted value on the hash table without replacing it with the new value to simplify our implementation.

\subsection{Cache Update on the GPU}
The proposed method is implemented in a GPU path tracer, which can assign more than one thread for different samples in a pixel.
Therefore, we use a lock-free algorithm using a compare-and-swap (CAS) instruction that avoids data races to insert the cacheable node's value into the cache.
For this CAS instruction, we encode the output value as a single 32-bit value, and then the second hash is packed with the encoded output value as a single 64-bit variable, which allows us to update the cache in a single CAS operation. 
Algorithm~\ref{alg:update} illustrates how to store the cacheable node's value in the cache using two different hashes with a 64-bit CAS instruction.

\section{Results}

We implemented our method of progressively caching materials using OpenCL\textsuperscript{\texttrademark}.
All the images in this paper are rendered on an AMD Radeon\textsuperscript{\texttrademark} RX 6900 XT GPU at $1920\times1080$ screen resolution.
Full-resolution images can be found in our supplemental material.

\begin{figure}[t]
  \centering
  \includegraphics[width=0.9\linewidth]{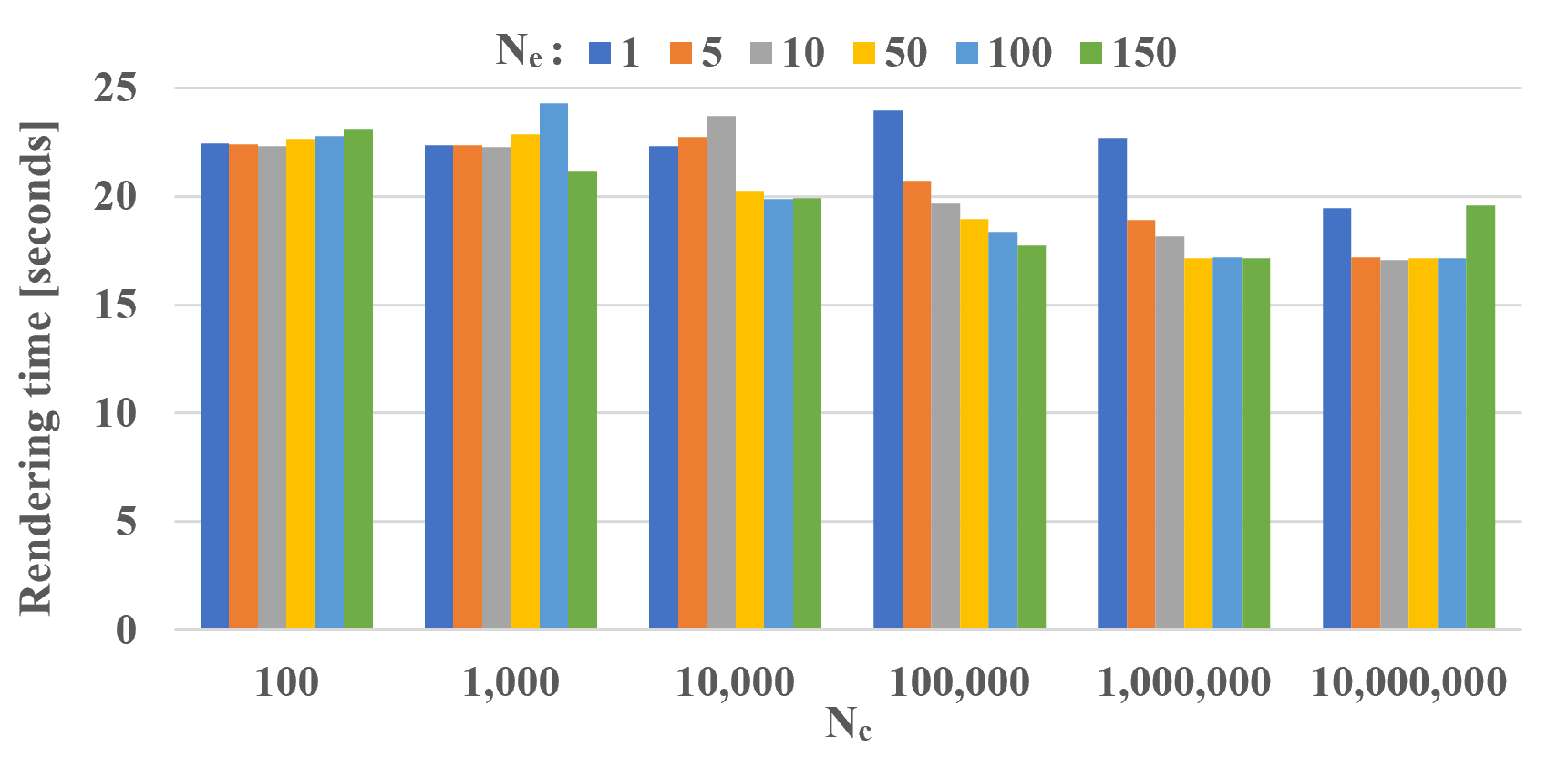}
  \caption{\label{fig:cacheSize}
           Cache size evaluation. $N_c$ and $N_e$ are the numbers of cells and entries on the cache, respectively.}
\end{figure}

\paragraph{Cache Size Evaluation}
We use the fixed-size cache in our implementation.
So, we investigate how the cache size affects the rendering performance.
We have two parameters, such as the numbers of cells $N_c$ and entries $N_e$, in the material cache.
Fig.~\ref{fig:cacheSize} shows the results of the performance evaluations with several combinations of $N_c$ and $N_e$ on the {\sc Classroom} scene with 32 samples per pixel.
In general, the larger $N_c$ improves the performance, while the larger $N_e$ usually shows better performance than the smaller $N_e$. 
However, the performance improvement hits its peak with some large size of the cache.
In this paper, we use $10^7 \times 10$ as the size of the cache that requires about $763$ MB of the memory, because it represents the best performance in this experiment.

\begin{figure}[t]
  \centering
  \includegraphics[width=0.75\linewidth]{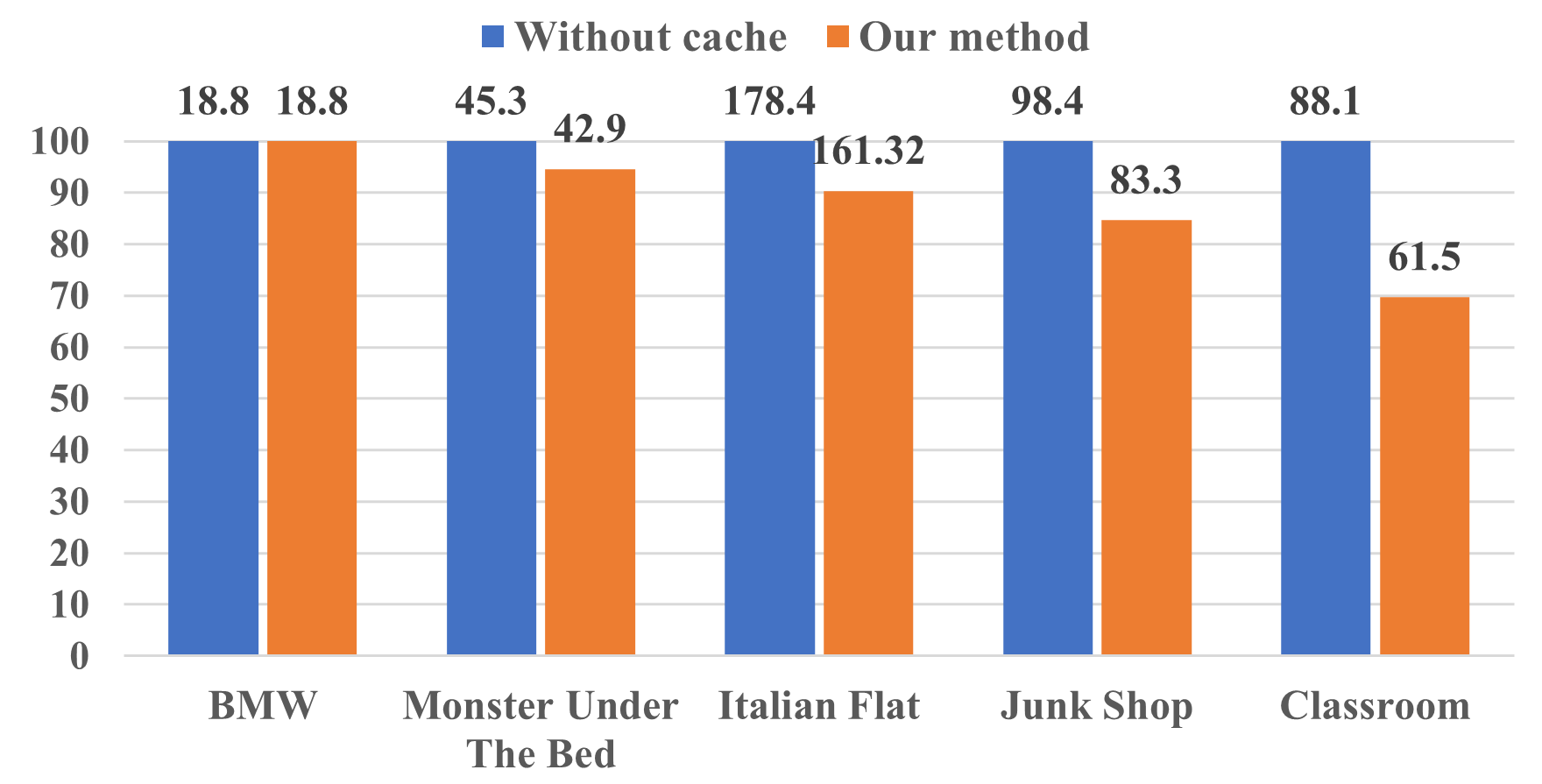}
  \caption{\label{fig:perfComp}
           Performance comparison of five scenes.}
\end{figure}

\paragraph{Performance Comparison}
Fig.~\ref{fig:perfComp} compares the performance on five scenes with 128 samples per pixel.
The vertical axis represents the percentage in the rendering time of our method as compared to that of the method without caching, taken as $100\%$.
Numbers on each bar show the actual rendering times in seconds.
On {\sc Classroom}, {\sc Junk Shop}, {\sc Italian Flat} and {\sc Monster Under The Bed} scenes, our method improves the performance by about $31\%$, $16\%$, $10\%$, and $5\%$ for each.
The {\sc Bmw} scene shows the worst case of our method with no gain in performance.
However, this result demonstrates that our method performs at least equally to the method without caching, because we can render the scene as it is, even if the material value cannot be stored or looked up in the cache.

\begin{figure}[t]
    \centering
    \begin{subfigure}[b]{0.32\linewidth}
        \centering
        \includegraphics[width=\linewidth]{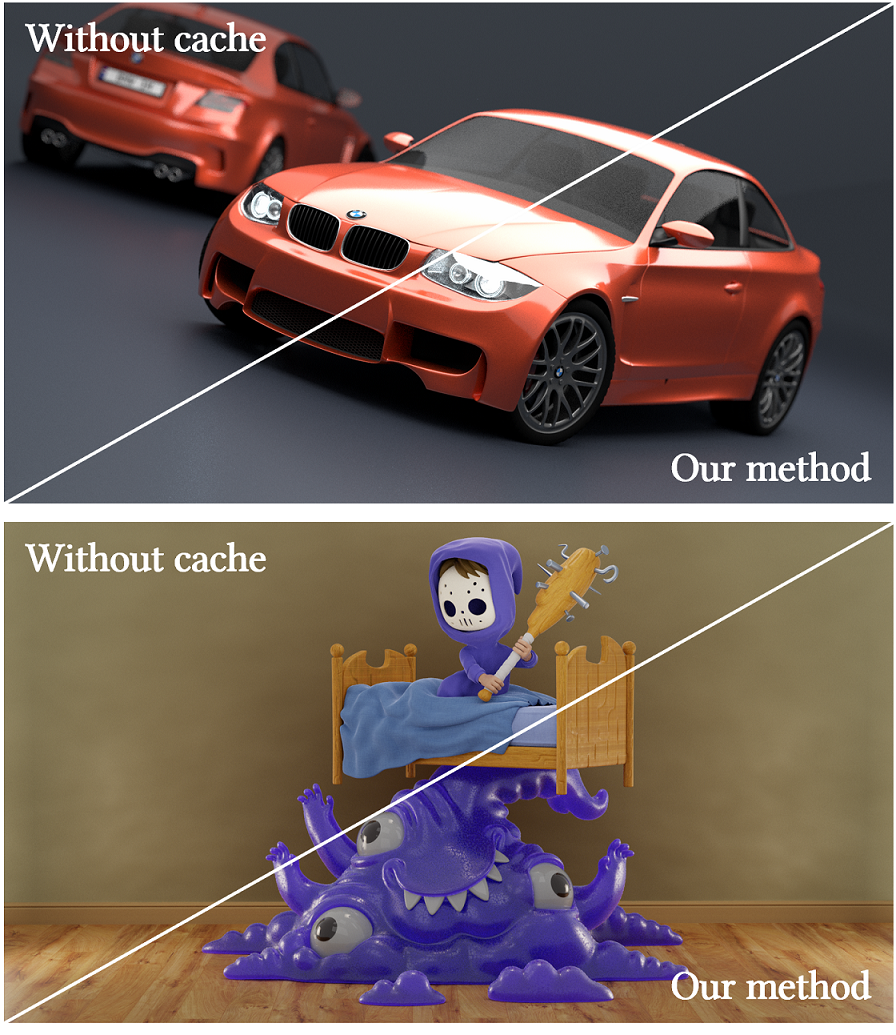}
        \caption{Rendered images}
    \end{subfigure}
    \hfill
    \begin{subfigure}[b]{0.32\linewidth}
        \centering
        \includegraphics[width=\linewidth]{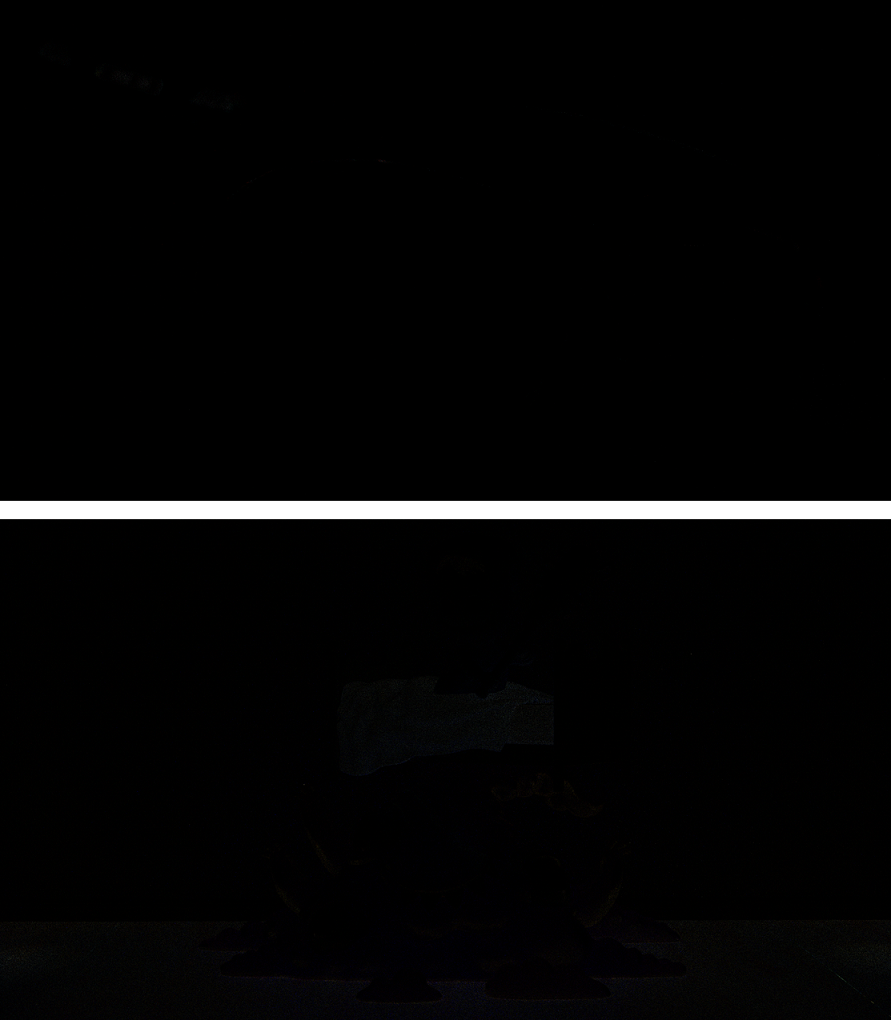}
        \caption{Image difference}
    \end{subfigure}
    \hfill
    \begin{subfigure}[b]{0.32\linewidth}
        \centering
        \includegraphics[width=\linewidth]{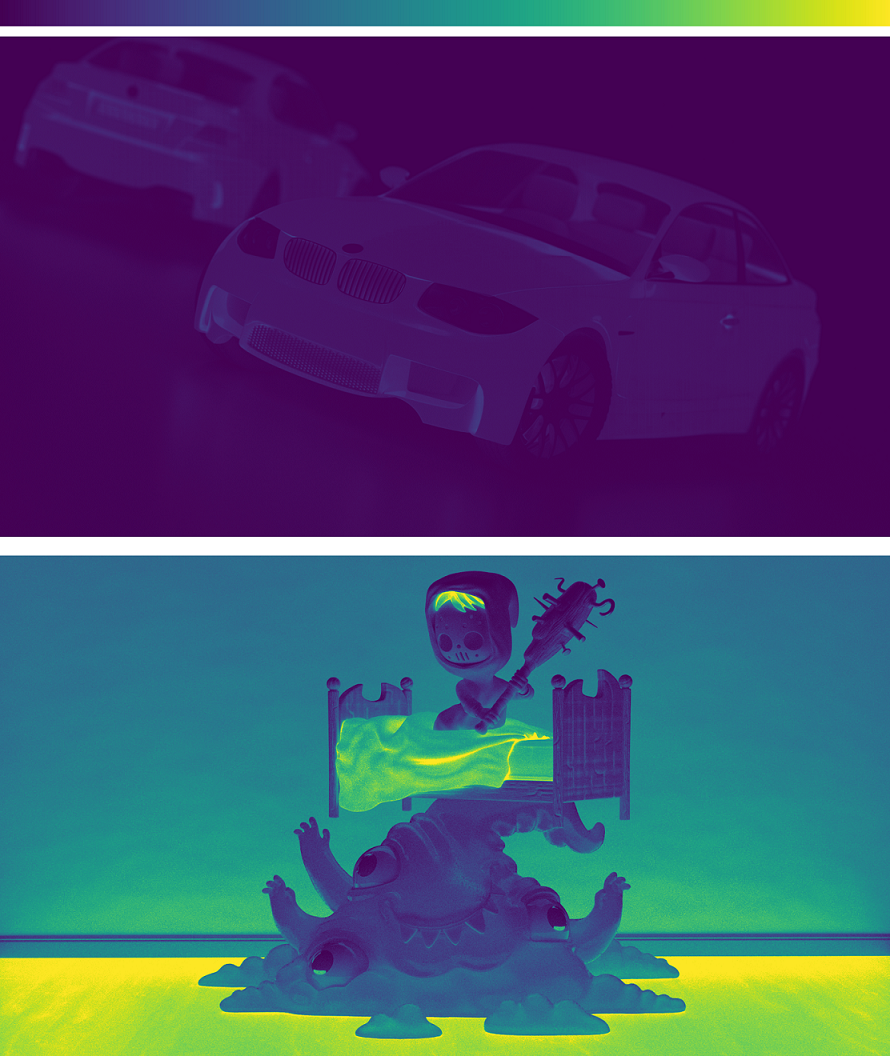}
        \caption{Number of nodes}
    \end{subfigure}
    \caption{Equal-sample comparisons (128 samples per pixel) with {\sc Bmw} and {\sc Monster Under The Bed} scenes. (a) Images rendered without the cache (top-left) and with our method (bottom-right). (b) and (c) are the same as Fig.~\ref{fig:teaser}c and Fig.~\ref{fig:teaser}d}
    \label{fig:compImages}
\end{figure}

Fig.~\ref{fig:teaser} and Fig.~\ref{fig:compImages} represent rendered results with the scenes on the experiment on Fig.~\ref{fig:perfComp}.
In Fig.~\ref{fig:teaser}a, Fig.~\ref{fig:teaser}b and Fig.~\ref{fig:compImages}a, we cannot see critical visual differences in the results, while the rendering time is reduced by $5 - 31\%$ except for the {\sc Bmw} scene.
Fig.~\ref{fig:teaser}c and Fig.~\ref{fig:compImages}b visualize those differences multiplied by 5.
As seen from these images, they do not show much discrepancy, although some surfaces with textures are slightly different due to the virtual texture's resolution.
As shown in Fig.~\ref{fig:virtualMip}, the virtual texture becomes more precise as the mipmap level becomes higher.
Therefore, by offsetting the mipmap level with a small value, we can obtain a bit more accurate result while it is a trade-off between the accuracy and the performance.
Also, Fig.~\ref{fig:teaser}d and Fig.~\ref{fig:compImages}c visualize the average number of material nodes per sample which we can find in the cache by lookup.
The color from black to yellow in the viridis colormap represents 0 to 20 times.
In the {\sc Classroom} scene, we find more nodes in the material cache over the whole scene compared to the {\sc Junk Shop} scene.
This can explain the difference in the performance improvement for these scenes.
However, comparing {\sc Classroom}, {\sc Italian Flat} and {\sc Monster Under The Bed} scenes, we note that the rendering time on the {\sc Classroom} scene is significantly reduced, while we can see more red in {\sc Italian Flat} and {\sc Monster Under The Bed} scenes.
This means we can cache the material nodes at the higher level in the network which contains more complex sub-graphs with our proposed method on the {\sc Classroom} scene.

\section{Conclusions and Future Work}
We presented an efficient approach to evaluate the material networks by progressively caching them without an overhead.
Our proposed method uses a fixed-size hash table to store and lookup the information about the cacheable nodes which can be effectively updated on the GPU.
The experiments show that our method can outperform significantly or at least perform equally to the method without caching.
The rendered images are slightly different, but the error does not appear an artifact.

In this paper, we demonstrated our method with the UV information as an additional descriptor of the cacheable node.
However, in the material networks, a material node can depend on other types of shading-point information which we can use as the descriptor.
We would like to extend our method to them in the future.
We expect supporting them will make the performance better because we can cache more material nodes in the network.
Our method uses fixed-size storage and keeps the first inserted values all the time, even if hash collisions happen.
This could be improved, because limited storage might be filled with data that is not reused often.
Thus, we also would like to investigate a more effective way of storing the material information in the cache, for example, by prioritizing directly visible materials over indirectly visible ones.

\begin{acks}
We thank Brian Savery at AMD for his inspiring study and valuable comments, Oleksandr Kupriyanchuk for proofreading.
We also thank the following people for their scenes used in our experiments: Christophe Seux ({\sc Classroom}), Alex Treviño and Anaïs Maamar ({\sc Junk Shop}), Flavio Della Tommas ({\sc Italian Flat}), Mike Pan ({\sc Bmw}) and Metin Seven and Blake Stevenson ({\sc Monster Under The Bed}). 
\end{acks}

\bibliographystyle{ACM-Reference-Format}
\bibliography{main}
\end{document}